# Field-induced water electrolysis switches an oxide semiconductor from an insulator to a metal

Hiromichi Ohta[1], Yukio Sato[2], Takeharu Kato[2], SungWng Kim[3], Kenji Nomura[3], Yuichi Ikuhara[2,4], & Hideo Hosono[3]

[1]*Graduate School of Engineering, Nagoya University, Furo-cho, Chikusa, Nagoya 464–8603, Japan & PRESTO, Japan Science and Technology Agency, Sanbancho, Tokyo 102–0075, Japan*
[2]*Nanostructures Research Laboratory, Japan Fine Ceramics Center, 2–4–1 Mutsuno, Atsuta, Nagoya 456–8587, Japan*
[3]*Frontier Research Center, Tokyo Institute of Technology, 4259 Nagatsuta, Midori, Yokohama 226–8503, Japan*
[4]*Institute of Engineering Innovation, The University of Tokyo, 2–11–16 Yayoi, Bunkyo, Tokyo 113–8656, Japan*

**Water is composed of two strong electro-chemically active agents, $H^+$ and $OH^-$ ions, for oxide semiconductors though water has never been utilized as an active electronic material. Here we demonstrate that water-infiltrated nanoporous glass electrically switches an oxide semiconductor from an insulator to metal. We fabricated the field effect transistor structure on an oxide semiconductor, $SrTiO_3$, using 100%-water-infiltrated nanoporous glass – amorphous $12CaO\cdot7Al_2O_3$ – as the gate insulator. For positive gate voltage, electron accumulation, water electrolysis and electrochemical reduction occur successively on the $SrTiO_3$ surface at room temperature, leading to the formation of a thin (~3 nm) metal layer with an extremely high electron concentration of $10^{15}$–$10^{16}$ cm$^{-2}$, which exhibits exotic thermoelectric behaviour. The electron activity of water as it infiltrates nanoporous glass may find many useful applications in electronics or energy storage.**

## Introduction

Water has never been utilized as an active electronic material though water has been widely applied for industrial uses such as coolant (radiators), solvent (batteries) and pressure medium (hydroelectric power generation). We have aimed to exploit the electrolysis of water.[1] Although the electrical conductivity of pure water is extremely low (~0.055 μScm$^{-1}$ at 25°C),[2] ionization of $H^+$ and $OH^-$ ions occurs when the bias voltage (greater than 1.23 V) is applied between two metallic electrodes immersed in water, as shown in Fig. 1a. The ions are then attracted to the cathode and anode. Finally, $H_2$ and $O_2$ gases are generated on the cathode and anode, respectively, via the electron transfer on the electrode surface. Since $H^+$ and $OH^-$ ions, which are strong reducing/oxidizing agents for most of oxide semiconductors,[3–8] are simultaneously produced in water electrolysis, one may expect that the electrical conductivity of an oxide semiconductor, which is strongly dependent on oxygen non-stoichiometry, can be modulated by utilizing the redox reaction between $H^+/OH^-$ ions and the oxide surface. However, water electrolysis and the redox reaction do not take place, because no electric field can be applied on the insulating oxide surface in the first place. Thus, the surface of the oxide must be conductive, as schematically shown in Fig. 1b.

We have found that water-infiltrated nanoporous glass overcomes this problem and switches an oxide semiconductor from an insulator to a metal. Electron accumulation, water electrolysis and redox take place successively on an oxide surface at room temperature (RT), leading to the formation of a thin metal layer on the oxide. We have fabricated a field effect transistor (FET) structure with source, drain and gate electrodes on an insulating oxide, using water-infiltrated nanoporous glass as the gate insulator. First, the insulating oxide surface becomes slightly conductive by applying the gate voltage due to electrostatic charge accumulation. Then, water electrolysis occurs between the gate and the oxide surface. Finally, a redox reaction takes place between $H^+/OH^-$ ions and the oxide surface. As a result, a thin metal layer is formed on an insulating oxide.

The key material to make the best use of the electron activity of water is nanoporous glass. We choose amorphous $12CaO \cdot 7Al_2O_3$ ($a$-C12A7) with nanoporous structure for this purpose. C12A7 is an abundant and environmentally benign material. Crystalline C12A7 becomes semiconductor[9] or metal[10, 11] when the clathrated free oxygen ions ($O^{2-}$), which are incorporated into the cage structure (~0.4 nm in diameter), are removed by chemical reduction treatment. On the other hand, "amorphous" C12A7 is a good electrical insulator because amorphous C12A7 does not have cage structure, we therefore ruled out the possibility of electrical conductivity of amorphous C12A7 film. Since C12A7 can be hydrated easily,[12] it is used commercially as a major constituent of aluminous cement. In 1987, Hosono and Abe found that a large amount of bubbles were generated in an $a$-C12A7 glass, when C12A7 melt was quenched under high oxygen pressure.[13] In the present study, we developed out a method to fabricate $a$-C12A7 film with nanoporous structure (CAN, hereafter) by pulsed laser deposition (Supplementary Fig. S1 and Methods section).

Here we show that water infiltrating CAN electrically switches an oxide semiconductor from an insulator to metal using the CAN-gated FET structure on a $SrTiO_3$ single crystal (Fig. 2a) as a proof of concept. Although $SrTiO_3$ is a wide bandgap (~3.2 eV) insulator, it becomes $n$-type conducting $SrTiO_{3-\delta}$ by appropriate reducing treatments.[5] Conducting $SrTiO_3$, especially when it is a two-dimensional conductor,[14,15] is of great importance as an active material for future electronic devices[16] because it has several potential advantages over conventional semiconductor-based electronic materials, such as transparency,[17] giant magnetoresistance,[18] and giant thermopower.[19] For positive gate voltage application to the CAN-gated $SrTiO_3$ FET, electron accumulation, water electrolysis and electrochemical reduction occur successively on the $SrTiO_3$ surface at room temperature, leading to the formation of a thin (~3 nm) metal layer with an extremely high electron concentration of $10^{15}$–$10^{16}$ cm$^{-2}$, which exhibits exotic thermoelectric behaviour.

## Results

### Water-infiltrated nanoporous glass, CAN.

The trilayer structure composed of Ti (20 nm)/CAN (200 nm)/SrTiO₃ is clearly observed in the cross-sectional transmission electron microscopy (TEM) image of the CAN-gated SrTiO₃ FET (Fig. 2b). Large amount of brighter contrasts (diameter ~10 nm) are seen throughout the CAN region. Furthermore, several dark contrasts with diameter less than 10 nm are observed in the Z-contrast, high-angle, annular dark field scanning transmission electron microscopy (HAADF-STEM) image of the CAN film (Fig. 2c). Judging from these TEM/STEM images, high density nanopores with a diameter of less than 10 nm are incorporated into the CAN film.

We subsequently performed thermal desorption spectrum (TDS) measurements of the CAN films to detect weakly bonded chemical species in the nanopores. Most of the desorbed species was $H_2O$ ($m/z$=18, where $m/z$ indicates the molecular mass to charge ratio) (Fig. 2d). The amount of $H_2O$ up to 400°C was estimated to be $1.4 \times 10^{22}$ cm$^{-3}$, which corresponds to 41%. The bulk density of the CAN film was ~2.1 gcm$^{-3}$, evaluated by grazing incidence X-ray reflectivity (see Supplementary Fig. S2), which corresponds to 72% of fully dense $a$-C12A7 (2.92 gcm$^{-3}$).[20] From these results, we judged that moisture in the air (humidity 40–50%) would infiltrate the CAN film most likely due to capillary effect, hence nanopores in the CAN film were filled with water.

### Field-induced water electrolysis.

We then measured the electron transport properties of the CAN-gated SrTiO₃ FET at RT. Figure 3 summarizes (a) drain current ($I_d$) vs. gate voltage ($V_g$) curves, (b) gate current ($I_g$) vs. $V_g$ curves, and (c) capacitance ($C$)-$V_g$ curve of the CAN-gated SrTiO₃ FET. Corresponding properties of the dense $a$-C12A7-gated SrTiO₃ FET[21] (d, e, f) were also measured for comparison. Both the channel length $L$ and the channel width $W$ were 400 μm. Dielectric permittivity ($\varepsilon_r$) of $a$-C12A7 was 12.[21] The gate voltage sweeps were performed in numerical order (*ex*. 1: −5 V → 0 V → +5 V → 0 V → −5 V). The $I_d$ vs. $V_g$ curves (a) show large anticlockwise hysteresis, indicating motion of mobile ions,[22] though very small clockwise hysteresis (~0.5 V) is seen in the dense $a$-C12A7-gated SrTiO₃ FET (d). Although very small, $I_g$ (~2 pA) is observed in the dense $a$-C12A7-gated SrTiO₃ FET (e), while the $I_g$ of the CAN-gated SrTiO₃ FET (b) increases exponentially up to 20 nA with $V_g$, suggesting that mobile ions transport electronic charge. The $C$ vs. $V_g$ curve (c) of the CAN-gated SrTiO₃ FET shows large anticlockwise hysteresis loop. The maximum $C$ is ~160 pF, ~76% of the value for the dense $a$-C12A7-gated SrTiO₃ FET [(f) ~210 pF], consistent with the fact that the volume fraction of dense $a$-C12A7 part in the CAN film is ~72 %.

We also observed a clear pinch-off and saturation in $I_d$ at low $V_g$ region (Supplementary Fig. S3), indicating that the operation of this FET conformed to standard FET theory at low gate voltage. Thus, we concluded that first the insulating SrTiO₃ surface became slightly conductive with gate voltage due to that electron charge was accumulated at the SrTiO₃ surface by pure electrostatic effect.

In order to clarify the role of water in the electrical transport properties, we measured the $I_d$ vs. $V_g$ characteristics of the CAN-gated SrTiO₃ FET at 0°C using a Peltier cooler, because H$^+$ and OH$^-$ ions cannot move through ice (Supplementary Fig. S4). The device does not show such large anticlockwise hysteresis at 0°C. Although the maximum $C$ increased from ~160 pF (25°C) to ~240 pF (0°C), $I_d$ at $V_g$ = +10 V decreased from ~5 μA (25°C) to ~1 μA (0°C), most likely due to the fact that water in the CAN acts as a simple gate dielectric at 0°C ($\varepsilon_{r\,25°C}$ ~78, $\varepsilon_{r\,0°C}$ ~88).[23] The field effect mobility ($\mu_{FE}$) value of the FET at 0°C is ~0.8 cm$^2$V$^{-1}$s$^{-1}$, comparable to that of the dense $a$-C12A7-gated SrTiO$_3$ FET (~2 cm$^2$V$^{-1}$s$^{-1}$), obtained from $\mu_{FE}$ = $g_m[(C·V_d)/(W·L)]^{-1}$, where $g_m$ was the transconductance $\partial I_d/\partial V_g$. We also measured Hall mobility ($\mu_{Hall}$) of a CAN-gated SrTiO$_3$ FET after several $V_g$ application (metallic state) and obtained $\mu_{Hall}$ values of 2.3–2.5 cm$^2$V$^{-1}$s$^{-1}$, which is approximately three times larger than the $\mu_{FE}$ value (~0.8 cm$^2$V$^{-1}$s$^{-1}$). These results clearly indicate that H$^+$ and OH$^-$ ions in the CAN are the main contributors to electron transport at RT.

**Insulator to metal switching.**

We then observed the electrochemical redox reaction of SrTiO$_3$. Figure 4a shows the changes of $I_d$ and $I_g$ during $V_g$ sweep from +10 to +40 V as a function of retention time. The $I_d$ gradually increases with retention time. The rate of $I_d$ increase grows with $V_g$. The $I_g$ also increases with $V_g$. The $I_d$ reaches ~1 mA at $V_g$ = +40 V. The $I_d$ decreases drastically when negative $V_g$ of −30 V is applied.

The sheet resistance ($R_{xx}$) at several points (A – E) was plotted as a function of ion density, which was obtained by $I_g·t$ (Fig. 4b). The observed values were close to the gray line, which is $R_{xx}$ = $(e·n_{xx}·\mu)^{-1}$, where $n_{xx}$ and $\mu$ are ion density and $\mu_{FE}$ (0.8 cm$^2$V$^{-1}$s$^{-1}$) was obtained as in Supplementary Fig. S4b, suggesting that carrier electrons were generated as a result of the electrochemical reduction of SrTiO$_3$. Although samples A – C exhibit insulating $R_{xx}$–$T$ behaviour, D and E exhibit metallic $R_{xx}$–$T$ behaviour (Fig. 4c).

**Exotic thermopower.**

We have also shown that the metal layer on the SrTiO$_3$ in the FET exhibits novel thermoelectric behaviour: V-shaped turnaround of $S$ is seen, although $R_{xx}$ decreases gradually as $V_g$ increases (Fig. 5). As we have reported previously, |$S$| value enhancement of electron doped SrTiO$_3$ can be observed when the conducting layer thickness is less than ~3 nm,[19] probably due to two-dimensional effect,[24] where the thermal de Broglie wavelength of conduction electrons in SrTiO$_3$ is ~6 nm.[25] We therefore expected that two-dimensional |$S$| can be observed at high $V_g$ because the layer thickness may become thinner.

When the $V_g$ was smaller than +26 V, |$S$| value gradually decreased with $V_g$. Similar |$S$| behaviour, which can be analyzed using simple the 3D electron diffusion theory, was also observed in the dense $a$-C12A7-gated SrTiO$_3$ FET as shown in Supplementary Fig. S5. Thus, we used the 3D electron diffusion theory[26] to analyze the layer thickness. The 3D electron concentration ($n_{3D}$) at $V_g$ = +16, +20, and +26 V were estimated to be ~1.5 × 10$^{19}$, ~1.5 × 10$^{20}$, and ~1.1 × 10$^{21}$ cm$^{-3}$, respectively. The $n_{xx}$ values at $V_g$ = +16, +20, and +26 V, which can be calculated using $R_{xx}$ and $\mu_{FE}$ (0.8 cm$^2$V$^{-1}$s$^{-1}$) are 1.4 × 10$^{13}$, 9.0 × 10$^{13}$, and 3.3 × 10$^{14}$ cm$^{-2}$, respectively, suggesting that the SrTiO$_{3−\delta}$ thickness

($\sim n_{xx}/n_{3D}$) are $\sim 9$, $\sim 6$, and $\sim 3$ nm, respectively, confirming the observed $S$ obeys simple 3D electron diffusion theory. On the contrary, when the $V_g$ greater than 26 V was applied, the $|S|$ value increased. The conducting layer thickness may be less than 3 nm, the observed upturn at $V_g > 26$ V is most likely due to two-dimensional effect.[19, 24]

**Discussion**

The present CAN-gated $SrTiO_3$ FET has several merits compared to established methods such as thermal reduction,[5] ion irradiation,[17, 27] thin film growth,[14, 15, 17, 18] and simple FET,[21, 28] which are often utilized to switch an insulating $SrTiO_3$ to a metal. This is because the current method allows a metal layer can be easily fabricated with extremely low energy. Firstly, required electricity is extremely small ($\sim 50$ $\mu Wcm^{-2}$ for "D" in Fig. 4) compared with that of thermal reduction ($\sim kW$) and ion implantation ($\sim Wcm^{-2}$). Secondly, an extremely thin ($\sim 3$ nm) metal layer, which exhibits exotic thermopower (Fig. 5), can be fabricated. Normally such thin layers must be fabricated through complicated vapour phase epitaxy methods, such as PLD and MBE. Thirdly, our CAN-gated $SrTiO_3$ FET exhibits a nonvolatile metal or highly conductive/insulator transition (see Figs. 3a, 4a, and Supplementary Fig. S6) because a reversible redox reaction is utilized in addition to the field effect.

Although several efficient gating methods using liquid electrolytes have been proposed very recently,[28–31] we would like to argue that the present water-infiltrated nanoporous glass "CAN" is truly superior to the liquid-based gate dielectrics. Liquid electrolytes including "gel" would be very useful to electrostatically accumulate carriers at the transistor channel by applying rather low $V_g$ (a few volts) using their huge capacitance. However, they would not be suitable for practical applications without sealing due to liquid leakage problem. Our "CAN" is a chemically stable rigid glassy solid with a higher decomposition voltage, showing excellent adhesion with chemically robust oxide such as $SrTiO_3$ surface and no liquid (water) leakage occurs. Carrier injection/discharge can be controlled by $V_g$ since Redox reaction, which occurs at the interface of $H^+$ or $OH^-$/semiconductor, is utilized for carrier injection/discharge, though rather high $V_g$ (several ten volts) should be required for Redox reaction. Further, we observed novel thermopower behaviour: V-shaped turn around, probably due to transition of electronic nature in the interface region from 3D to 2D nature, since the FET could be operated at high $V_g$. These clearly indicate the effectiveness of our "CAN".

In summary, we have demonstrated that water, when it infiltrates nanoporous glass, can switch an insulating oxide to metal. As an example, we have built a FET on an insulating oxide, $SrTiO_3$, using water-infiltrated nanoporous glass, amorphous $12CaO \cdot 7Al_2O_3$ with nanoporous structure, CAN, as the gate insulator. First, the insulating $SrTiO_3$ surface became slightly conductive with gate voltage due to electrostatic charge accumulation. Then, $H^+/OH^-$ ions were generated due to water electrolysis occurring between the gate and the $SrTiO_3$ surface. Subsequently, a redox reaction took place between $H^+/OH^-$ ions and the $SrTiO_3$ surface. As a result, a thin metal ($\sim 3$ nm) layer with extremely high electron concentration of $10^{15}$–$10^{16}$ $cm^{-2}$ was formed on the insulating $SrTiO_3$. The electron activity of water as it infiltrates nanoporous glass "CAN" may find many useful applications in electronics or energy storage.

**Methods**

**Fabrication of the CAN-gated SrTiO₃ FET**

The FET structures (Fig. 2a) were fabricated on the (001)-face of $SrTiO_3$ single crystal plates (10 mm × 10 mm × 0.5 mm, SHINKOSHA Co. Japan), treated in NH₄F-buffered HF (BHF) solution.[32] First, 20-nm-thick metallic Ti films, used as the source and drain electrodes, were deposited through a stencil mask by electron beam (EB) evaporation (base pressure ~$10^{-4}$ Pa, no substrate heating/cooling) onto the $SrTiO_3$ plate. Ohmic contact between the Ti and $SrTiO_3$ surface was confirmed by conventional *I–V* characteristics (output characteristics) as shown in Supplementary Fig. S3b. Then, a 200-nm-thick CAN film was deposited through a stencil mask by pulsed laser deposition (PLD, KrF excimer laser, fluence ~3 $Jcm^{-2}pulse^{-1}$) at RT using dense polycrystalline C12A7 ceramic as target. During the CAN deposition, the oxygen pressure in the deposition chamber was kept at 5 Pa. Finally, a 20-nm-thick metallic Ti film, used as the gate electrode, was deposited through a stencil mask by EB evaporation.

**Analysis of the CAN films**

The bulk density and thickness of the CAN films were evaluated by grazing incidence X-ray reflectivity (GIXR, ATX-G, Rigaku Co. Japan). Microstructures of the CAN films were observed by using transmission electron microscopy (TEM, JEOL JEM-2010, acceleration voltage = 200 kV) and aberration-corrected scanning transmission electron microscope (STEM, JEOL JEM-2100F, acceleration voltage = 200 kV). Water concentration in the CAN films was analyzed by thermal desorption spectrum (TDS) measurements (TDS1200, ESCO Ltd. Japan). The TDS measurements were carried out in a vacuum chamber with the background pressure of ~$10^{-7}$ Pa at temperatures varied from 60 to 700°C at a heating rate of 60°C min⁻¹.

**Electron transport properties**

Electrical properties of the FETs were measured by using a semiconductor device analyzer (B1500A, Agilent). The capacitance of the CAN layer on the FETs was measured using an LCR meter (4284A, Agilent). The thermopower (*S*) values were measured using two Peltier devices under the FET to give a temperature difference between the source and drain electrodes (Supplementary Fig. S7). Two thermocouples (K-type) located at both ends of the channel were used for monitoring the temperature difference (Δ*T*, 0–5 K). *S* values were measured after the each *V*_g sweep (*ex.* *V*_g application 0 V → +16 V → 0 V → *S* measurements).

**Acknowledgements**

We thank D. Kurita, A. Yoshikawa, T. Mizuno for technical assistance and R. Asahi for discussions. H.O. is supported by MEXT (22360271, 22015009). Y.S. is supported by Research Fellowships of JSPS for young scientists. The Research at Tokyo Tech. is supported by JSPS–FIRST Program.

**Contributions**
H.O. performed the sample fabrication, measurements and data analysis. Y.S., T.K., and Y.I. performed TEM/STEM analyses. S.K. supplied dense C12A7 ceramics. S. K., K.N. and H.H. contributed to water analyses. All authors discussed the results and commented on the manuscript. H.O. planned and supervised the project.

**Competing financial interests**

**Corresponding author**
Correspondence and requests for materials should be addressed to: H. Ohta (h-ohta@apchem.nagoya-u.ac.jp)

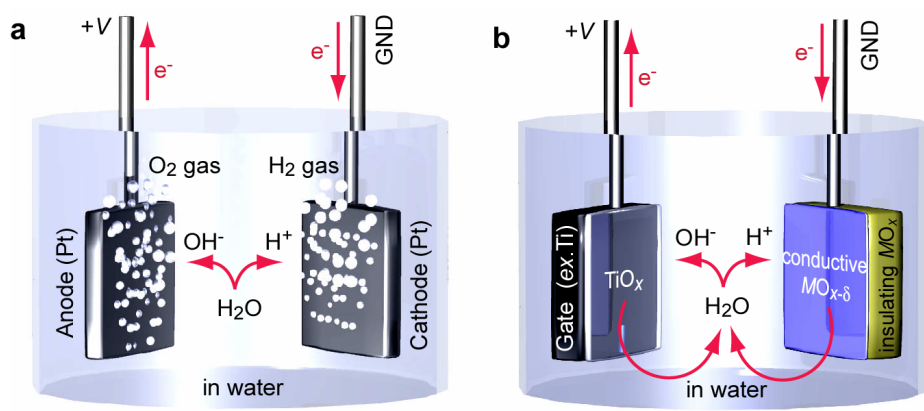

**Figure 1: Field-induced water electrolysis switches an oxide semiconductor from an insulator to a metal.** (**a**) Simple water electrolysis with two Pt electrodes as the cathode and anode immersed in water. $H^+$ and $OH^-$ ions, which are generated by the electrolysis, become $H_2$ and $O_2$ gases on the anode and cathode, respectively. (**b**) Water electrolysis with an insulating oxide $MO_x$, with a slightly conductive surface $MO_{x-\delta}$. Similar to (a), $H^+/OH^-$ ions are attracted to the $MO_{x-\delta}$, leading to the redox reaction between $H^+/OH^-$ ions and the $MO_{x-\delta}$ surface.

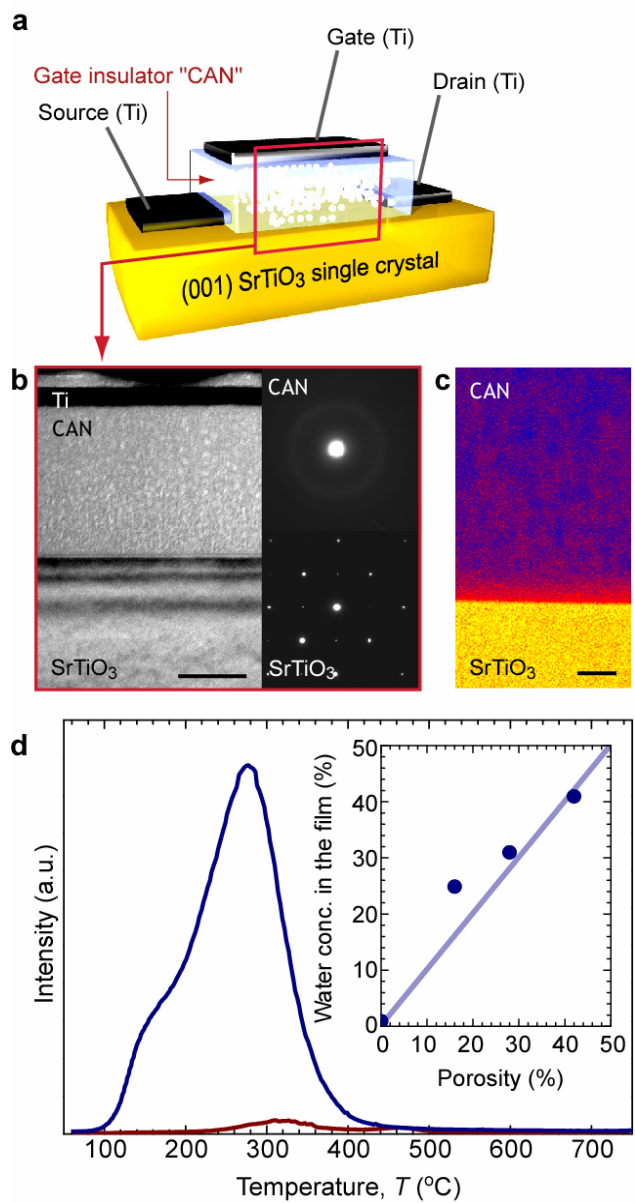

**Figure 2: The CAN-gated SrTiO₃ FET.** (**a**) A schematic illustration of the CAN-gated SrTiO₃ FET. (**b**) Cross-sectional TEM image of the CAN-gated SrTiO₃ FET (left panel). Scale bar is 100 nm. Trilayer structure composed of Ti/CAN/SrTiO₃ is observed. Large amount of light spots are seen in the whole CAN region. Broad halo is observed in the selected area electron diffraction pattern of CAN, and diffraction pattern from SrTiO₃ single crystal is also shown below (right panel). (**c**) Z-contrast HAADF-STEM image of the CAN/SrTiO₃ interface. Scale bar is 20 nm. Nanopores with diameter less than 10 nm appear dark. (**d**) TDS spectrum of water ($m/z$ = 18 $H_2O$) in the CAN film (blue). The amount of $H_2O$ up to 400°C was estimated to be 1.4 × 10²² cm⁻³ (=0.41 gcm⁻³, ~41%). TDS spectrum in the dense $a$-C12A7 film is shown for comparison (red, 0.009 gcm⁻³). The water concentration increases monotonically with an increase of the porosity of the CAN films (the inset).

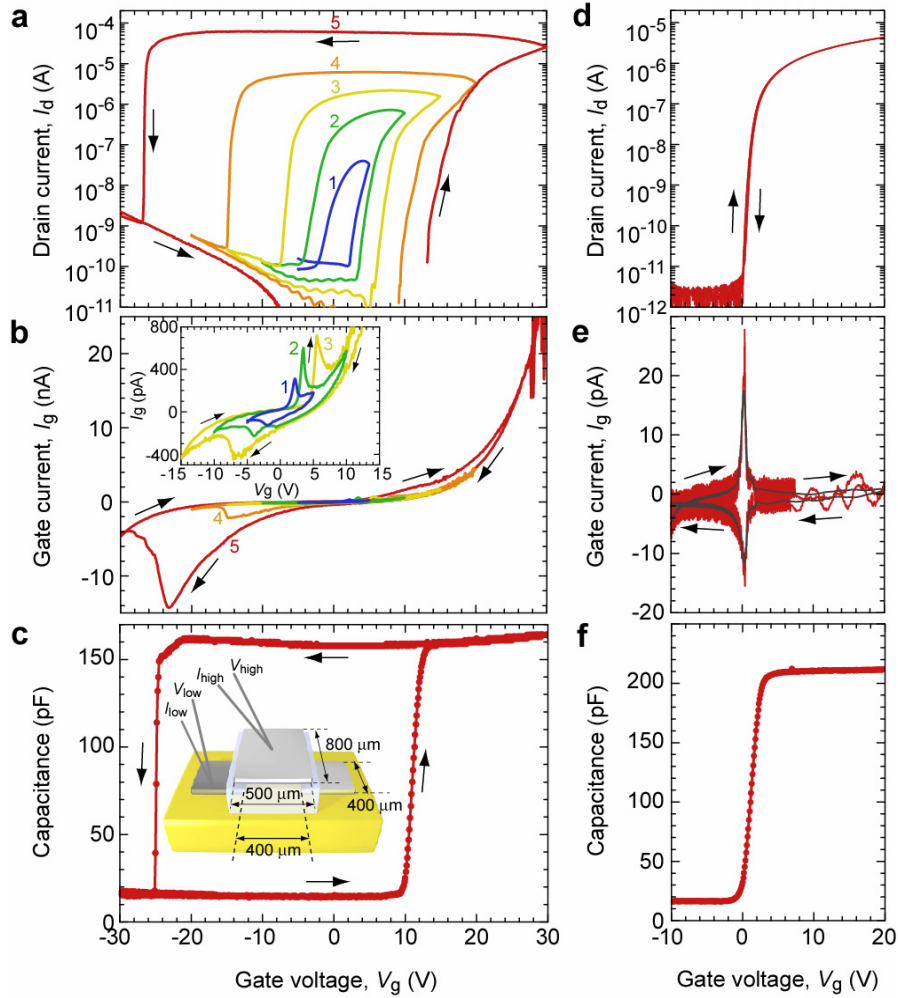

**Figure 3: Electron transport properties of the CAN-gated SrTiO₃ FET at RT.**
(**a**) $I_d$ vs. $V_g$, (**b**) $I_g$ vs. $V_g$, and (**c**) $C$-$V_g$. $I_d$ vs. $V_g$ (**d**), $I_g$ vs. $V_g$ (**e**), and $C$-$V_g$ curve
(**f**) of the dense $a$-C12A7-gated SrTiO₃ FET are also shown for comparison. Both
the channel length $L$ and the channel width $W$ are 400 μm. The gate voltage
sweeps were performed in numerical order (*ex.* 1: −5 V → 0 V → +5 V → 0 V →
−5 V). (**a**) $I_d$ vs. $V_g$ curves ($V_d$ = +2 V) show large anticlockwise hysteresis,
although very small clockwise hysteresis is seen in the dense one (**d**, $V_d$ = +1 V).
(**b**) $I_g$ increases exponentially up to 20 nA with $V_g$, which is ~$10^4$ greater than that
of the dense $a$-C12A7-gated SrTiO₃ FET (**e**, red: observed, gray: smoothed). (**c**)
The $C$ vs. $V_g$ curve shows a large anticlockwise hysteresis loop. The maximum $C$
(frequency: 20 Hz) of the nanoporous $a$-C12A7-gated SrTiO₃ FET is ~160 pF,
~76% of that of the dense $a$-C12A7-gated SrTiO₃ FET [(**f**) ~210 pF].

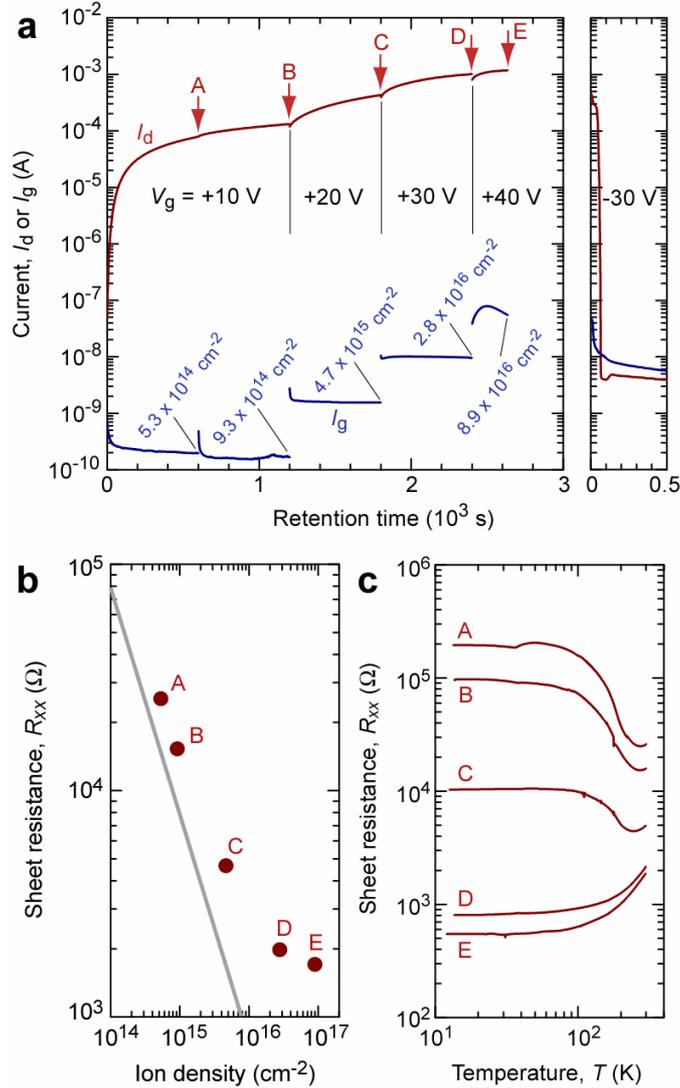

**Figure 4: A redox reaction switches an insulating SrTiO$_3$ to metal.** (a) Retention time dependences of $I_d$ and $I_g$ for the CAN-gated SrTiO$_3$ FET at several $V_g$ at RT ($V_d$ = +2 V). The $I_d$ increases gradually with retention time at constant $V_g$. Ion density, which is the retention time integral of $I_g$, reaches ~9 × 10$^{16}$ cm$^{-2}$ when $V_g$ = +40 V is applied. (b) $R_{xx}$ vs. ion density for the state A − E marked in (a) at RT. The gray line (slope = −1) indicates $R_{xx}$ = ($e \cdot n_{xx} \cdot \mu$)$^{-1}$, where $n_{xx}$ and $\mu$ are ion density and $\mu_{FE}$ (0.8 cm$^2$V$^{-1}$s$^{-1}$) obtained in Fig. 4b. (c) Temperature dependence of $R_{xx}$ for the state A − E marked in Fig. 4a.

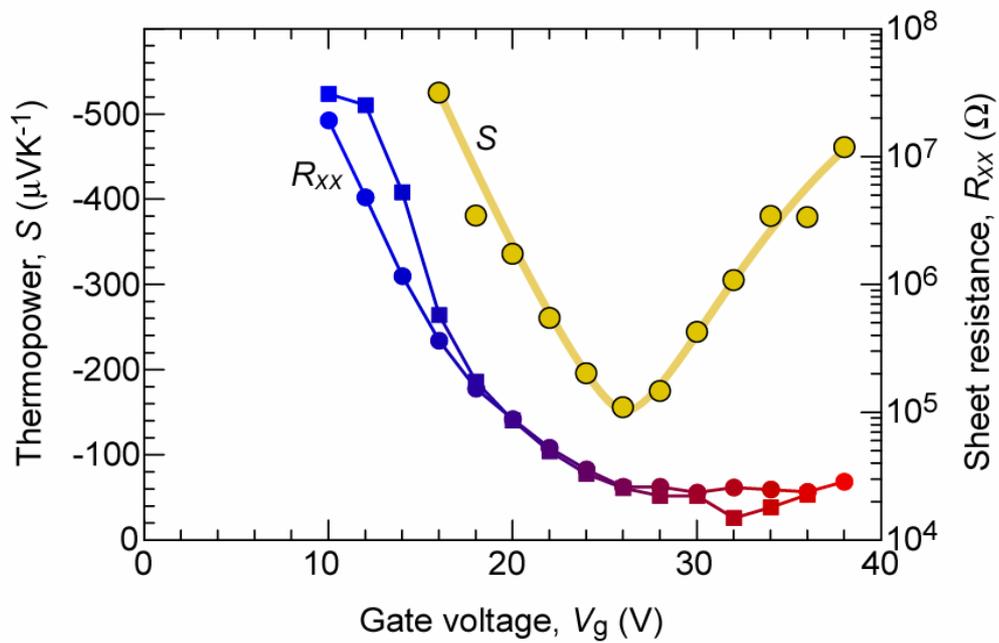

**Figure 5: Thermopower and sheet resistance as a function of applied *V*g.**
Novel thermopower behaviour: V-shaped turnaround of *S* is seen, though *Rxx*
(circle: before *S* measurement, square: after *S* measurement) decreases
gradually as *V*g increases, probably due to transition of electronic nature in the
interface region from 3D to 2D nature.

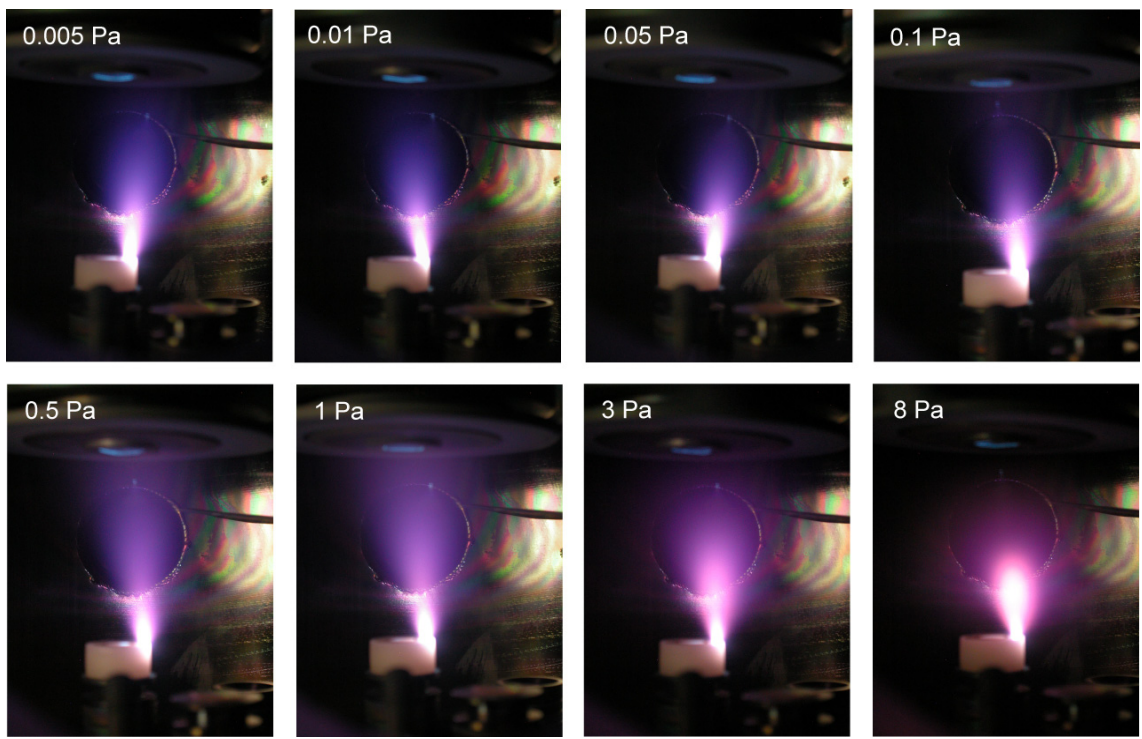

**Supplementary Figure S1: The ablation plumes of C12A7 under several oxygen pressures.** Amorphous $12CaO \cdot 7Al_2O_3$ (*a*-C12A7) films were deposited by a pulsed laser deposition (PLD, KrF excimer laser, fluence ~3 $Jcm^{-2}pulse^{-1}$) at room temperature using dense polycrystalline C12A7 ceramic as target. During the C12A7 deposition, the oxygen pressure in the deposition chamber was varied from 0.005 to 8 Pa. When the oxygen pressure exceeds 0.1 Pa, the plume becomes large.

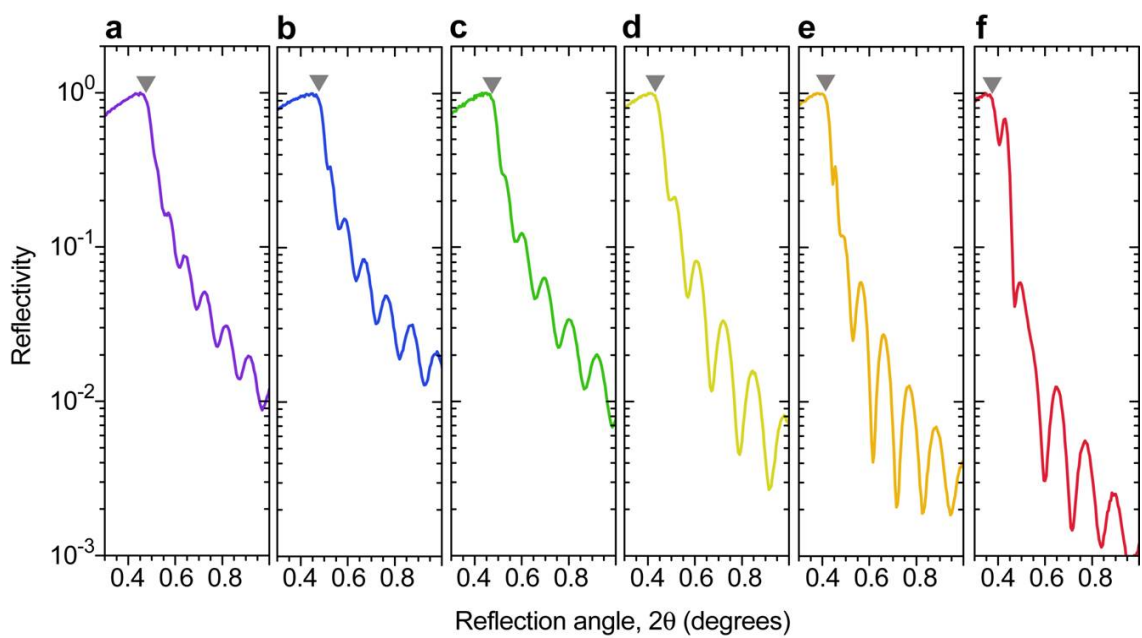

**Supplementary Figure S2: The GIXR patterns of the C12A7 films deposited at various oxygen pressures.** Bulk density and thickness of the C12A7 films were evaluated by grazing incidence X-ray reflectivity (GIXR, ATX-G, Rigaku Co. Japan). [Oxygen pressure, (**a**) 0.1 Pa, (**b**) 0.3 Pa, (**c**) 1 Pa, (**d**) 3 Pa, (**e**) 5 Pa, (**f**) 8 Pa] [Bulk density, (**a**) 2.9 cm$^{-3}$, (**b**) 2.9 cm$^{-3}$, (**c**) 2.9 cm$^{-3}$, (**d**) 2.4 cm$^{-3}$, (**e**) 2.1 cm$^{-3}$, (**f**) 1.8 cm$^{-3}$] At low oxygen pressures, the bulk density is ~2.9 gcm$^{-3}$, corresponding well with the full density (2.92 gcm$^{-3}$). When the oxygen pressure exceed 1 Pa, the bulk density decreases gradually with oxygen pressure.

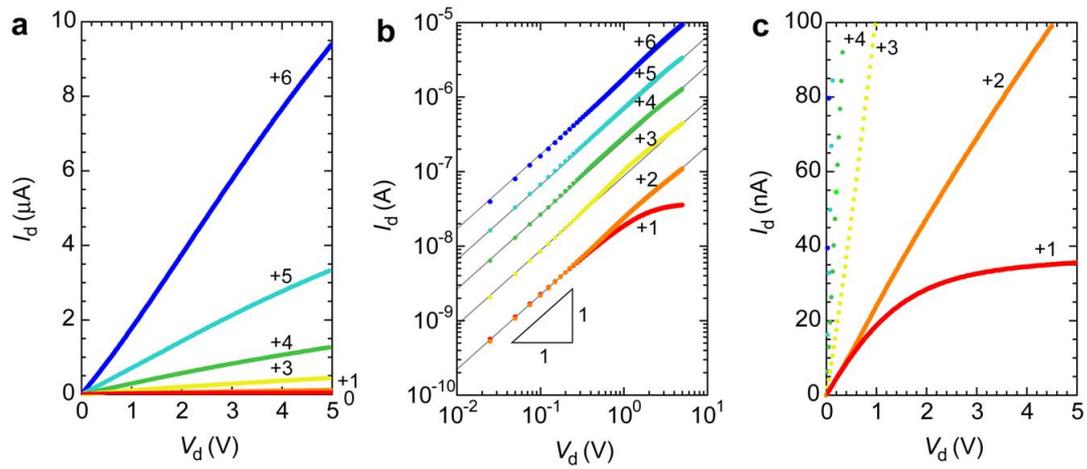

**Supplementary Figure S3: Output characteristics of the CAN-gated SrTiO$_3$ FET.** (**a**) Linear scaled $I_d$ vs. $V_d$ plots. Effective gate voltages ($V_g$–$V_{gth}$) were varied from +1 to +6 V. (**b**) Logarithmic scaled $I_d$ vs. $V_d$ plots. Linear $I_d$–$V_d$ characteristics (**b**) clearly indicate Ohmic contact between the Ti and SrTiO$_3$ surface. (**c**) Magnified $I_d$ vs. $V_d$ plots (linear scaled). A clear pinch-off and saturation in $I_d$ are observed at $V_g$–$V_{gth}$ = +1 V (**c**), indicating that the operation of this FET conformed to standard FET theory at low gate voltage.

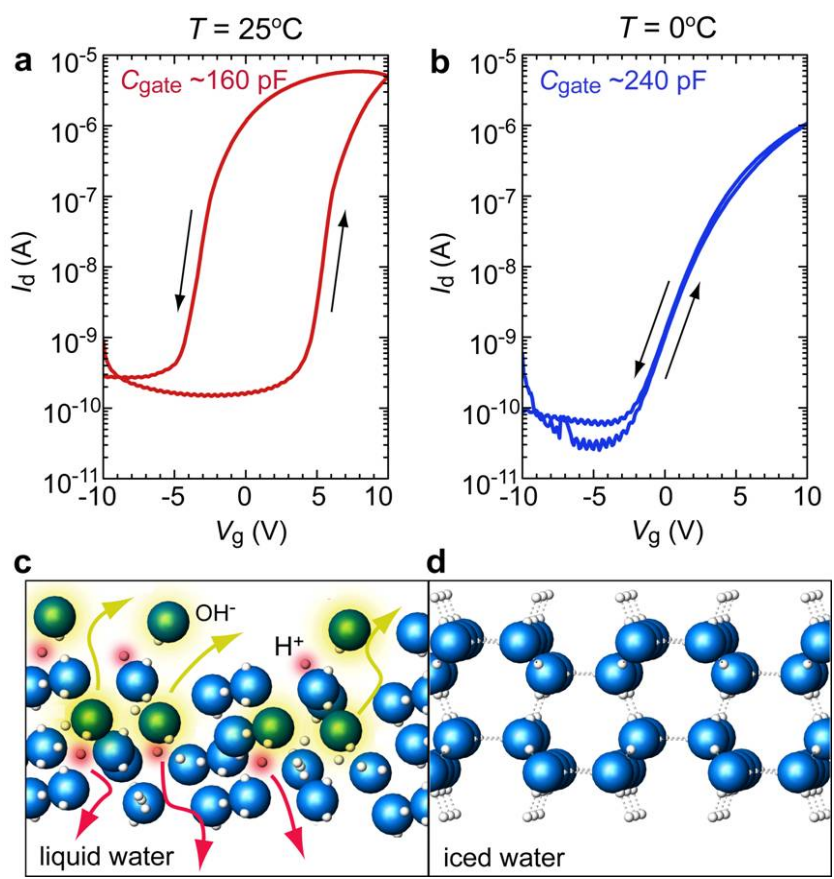

**Supplementary Figure S4: Role of liquid water in the CAN film.** $I_d$ vs. $V_g$ curves obtained at (**a**) 25°C and (**b**) 0°C. Schematic illustration of (**c**) liquid water and (**d**) iced water. Ion diffusivity of $H^+$ and $OH^-$ in liquid water (**c**) is incredibly faster than that in iced water. Both nanoporous C12A7 glass framework ($\varepsilon_r =$ 12) and neat water ($\varepsilon_r = $ 78) would play as gate dielectrics at low gate voltage (less than threshold voltage of water electrolysis, 1.23 V). Anticlockwise hystelysis is disappeared when the FET is cooled to the freezing point though the $C_{gate}$ increases. Iced water may play as gate dielectric (dielectric permittivity of iced water at 0°C, $\varepsilon_r$ ~88).

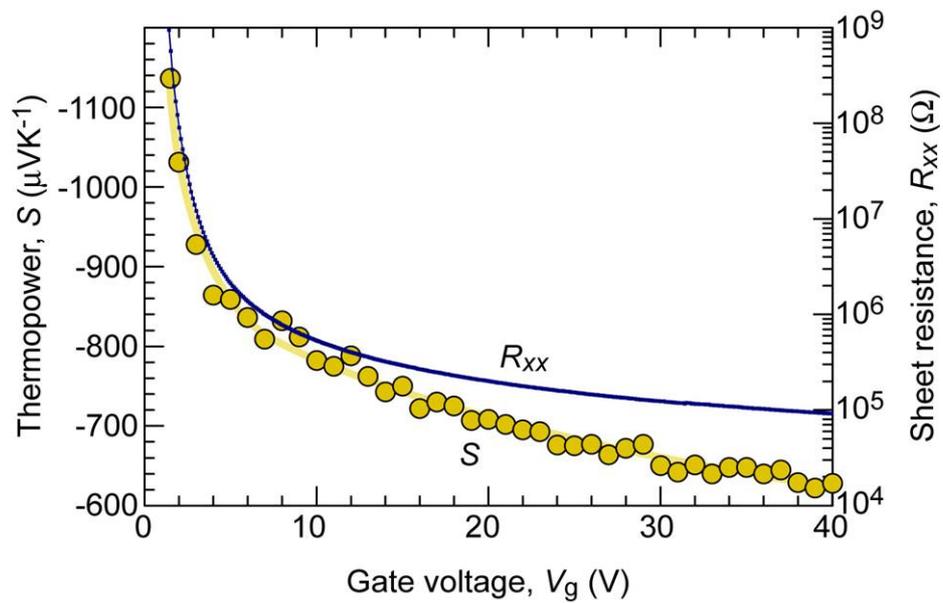

**Supplementary Figure S5: Field-modulation of *S* for the dense *a*-C12A7-gated SrTiO₃ FET channel at RT.** The |*S*|-value gradually decreases from 1140 to 620 µVK⁻¹, which corresponds to an increase of the $n_{3D}$ from ~1 × $10^{16}$ to ~4 × $10^{18}$ cm⁻³ due to the fact that electron carriers are accumulated by positive $V_g$ (up to +40 V).

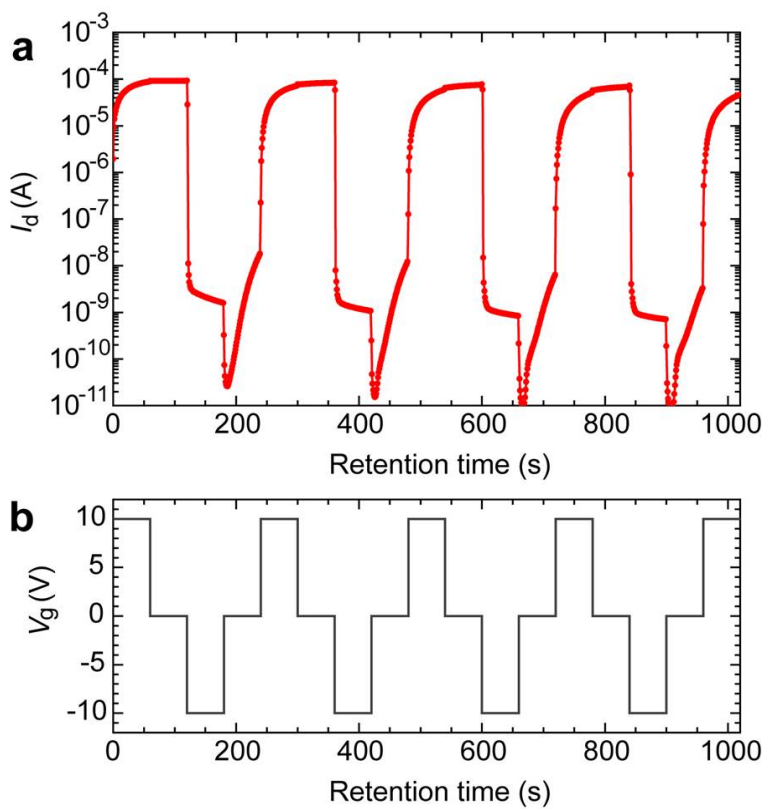

**Supplementary Figure S6: Nonvolatile memory behaviour of the CAN-gated SrTiO$_3$ FET at RT.** (**a**) Retention time dependence of $I_d$ ($V_d$ = +2 V, $L/W$ = 800/400 μm) during $V_g$ application as shown in (**b**) (−10 ~ +10 V). Nonvolatile behavior is clearly observed from highly conductive ($I_d$ ~10$^{-4}$ A) to insulator ($I_d$ ~10$^{-9}$ A). The switching speed of the device is very slow as compared to that of pure electrostatic FETs. This also indicates that Redox reaction dominates the device switching. Especially, reduction of the SrTiO$_3$ or oxidation of the Ti is slower than oxidation of the SrTiO$_{3-\delta}$ or reduction of the TiO$_x$. The on current gradually decreased with the number of switching most likely due to the fact that the gate Ti film is gradually oxidized during the positive $V_g$ application as illustrated in Fig. 1b. Further, the source Ti film is gradually oxidized during the negative $V_g$ application. We therefore consider that oxidation of Ti film may prevent effective $V_g$ application.

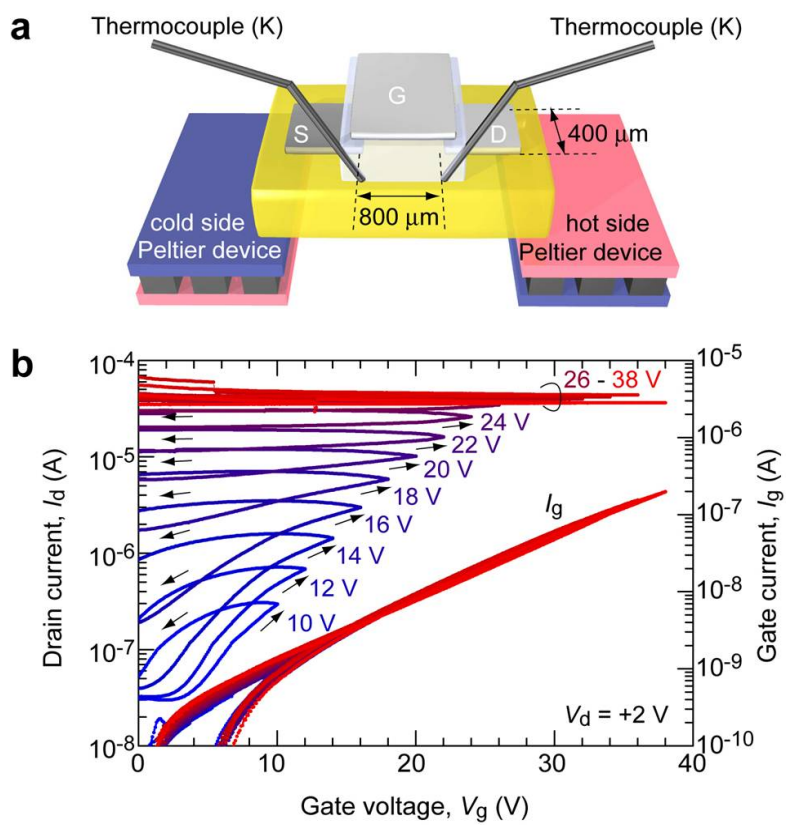

**Supplementary Figure S7: Thermopower measurements of the CAN-gated SrTiO₃ FET.** (**a**) Schematic illustration of the measurement configuration for thermopower $S$ at RT. Two Peltier devices under the FET are used to give a temperature difference between the source and drain electrodes. Two thermocouples (type K) located both ends of the channel are used for monitoring the temperature difference ($\Delta T$, 0~5 K). (**b**) $I_d$ vs. $V_g$ curves of the CAN-gated SrTiO₃ FET. $S$ values were measured after the each $V_g$ sweeping (*ex.* $V_g$ application 0 V → +16 V → 0 V, then $S$ measurement).